\documentclass[aps,twocolumn,showpacs,preprintnumbers,amsmath,amssymb]{revtex4}

\usepackage{graphicx}
\usepackage{dcolumn}
\usepackage{bm}
\usepackage{eurosym}
\usepackage{lscape}

\begin{document}

\title{A network approach for the scientific collaboration in the European Framework Programs}

\author{Antonios Garas}
 \altaffiliation[Also at ]{Department of Physics, University of Thessaloniki, 54124 Thessaloniki Greece.}
 \email{agara@physics.auth.gr}
\author{Panos Argyrakis}%
\affiliation{%
Department of Physics, University of Thessaloniki, 54124 Thessaloniki Greece.}%
 \email{panos@physics.auth.gr}

\date{\today}

\begin{abstract}
We construct the networks of collaboration between partners for projects carried out with the support of European Commission Framework Programs FP5 and FP6. We analyze in detail these networks, not only in terms of total number of projects, but also for the different tools employed, the different geographical partitions, and the different thematic areas. For all cases we find a scale free behavior, as expected for such social networks, and also reported in the literature. In comparing FP5 to FP6, we show that despite a decrease in the number of signed contracts, and the total number of unique partners, there is an increase in the average number of collaborative partners per institution. Furthermore, we establish a measure for the central role (hub) for each country, by using the Minimum Spanning Tree (MST), which we construct in detail for each thematic area (e.g. Informatics, Nanoscience, Life Sciences, etc.). The importance of these network hubs is highlighted, as this information can be used by policy planners in designing future research plans regarding the distribution of available funds.
\end{abstract}


\maketitle

\section{Introduction}

Complex network theory~\cite{bb:AlbertBarabasi} is a fast emerging field in physics research. Tools of complex network analysis were implemented for the analysis of a variety of systems in many fields, spanning a very wide spectrum of applications. Examples are the Internet~\cite{bb:Faloutsos,bb:ShaiCarmi}, the World Wide Web~\cite{bb:AlbertJeongBarabasi}, communication networks~\cite{bb:PastorVespignani}, food webs~\cite{bb:Garlaschelli}, sexual contact networks~\cite{bb:Liljeros}, scientific collaboration networks~\cite{bb:NewmanPnas}, economic networks~\cite{bb:Mantegna,bb:Tumminello,bb:OnnellaKaskiKertesz,bb:GarasArgyrakis,bb:GarasArgyrakisHavlin}, and many more. These tools use a novel approach to deal with nature and society and they can lead beyond usual statistical analysis, uncovering properties not palpable by classical means.

One important network is formed by the collaboration of scientific and industrial
institutions in specific projects of a finite duration. This typically involves a group of universities, research centers, and private companies which join their efforts in order to solve a particular problem. Typically, such projects originate by the initiative of the scientists involved, who respond to the specific calls for proposals by the funding agencies. It would thus be of interest to investigate the patterns by which scientists connect among themselves, which in turn would provide valuable information to the government and to the policy makers when looking to formulate their strategy for crucial everyday problems in environment, energy, communications, etc.

The European Commission (EC) has established the European Research Area (ERA), which has been given the responsibility to guide Europe to the highest quality research that can be performed in the world today. To achieve its goal the ERA sponsors the so called Framework Programs (FP), which are large funding programs for a certain length of time, with a fixed budget, and which are expected to produce tangible results. In order to ensure the diffusion of knowledge and funding resources with the FPs, international collaborations are strongly encouraged. There has been a total of seven FPs up to now. The latest FP is the 7$^{\textrm{th}}$ Framework Program (FP7) which runs in the period 2007-2013 with a total budget of $\sim$ \EUR{51} billion. This program is comprised of different funding schemes with varying scopes, targets, and different number of participants. Among such schemes are: the STREP scheme which focuses on a specific target problem in a consortium of about 8-10 partners, the Integrated Projects (IPs) which are larger collaborations of about 20 - 25 partners, dealing with problems in a broader area but with a unifying theme, the Marie-Curie Actions (MCA) that are focused on the exchange of graduate students and postdoctoral scholars, usually between 2 partners, and several more schemes.

In this work we analyze the collaboration network for the previous two FPs which have been now concluded, the FP5 and the FP6, and were carried out in the periods 1996-2001 and 2002-2006, respectively. We use the entire dataset of all projects that were approved, funded, and carried out in the FPs. The information for such data can be obtained from CORDIS~\cite{bb:CORDIS}. To give an overview of the size of the projects, we simply mention that in FP5 there were 16558 contracts that were carried out by 84267 partners (but only 27219 unique partners) from 147 countries, while in FP6 there were 8861 contracts that were carried out by 69237 partners (but only 19984 unique partners) from 154 countries. In a recent paper~\cite{bb:MendesFP5} Almendaral el al. studied some properties of the FP5 network. They found that the network is scale-free with an accelerated growth, and due to the hierarchical modularity property, it has a self similar structure. They also found that the network features assortative mixing, which means that collaborations among participants of similar size appear easier, and it possesses the property of small world. In the present work we start by doing a comparative analysis between the FP5 and FP6 networks, but eventually we will use tools of graph theory~\cite{bb:West} to analyze in more detail the FP6 network.

\begin{figure}
\resizebox{1.0\columnwidth}{!}{%
\includegraphics{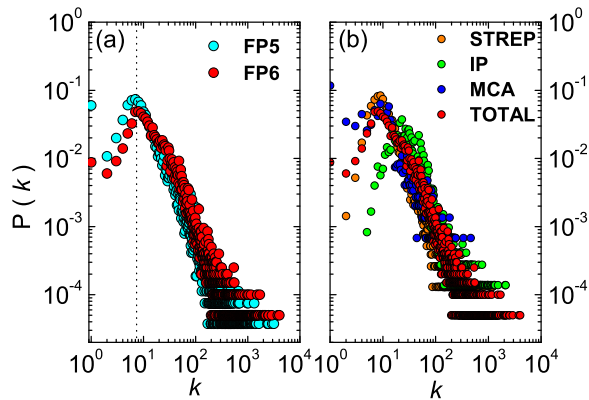}
}
\caption{(a) Probability distribution of the degrees for the FP5 and FP6 collaboration networks.
(b) Probability distribution of the degrees of the collaboration networks for different instruments of FP6.
}
\label{fig:1}
\end{figure}

\begin{figure}
\resizebox{0.95\columnwidth}{!}{%
\includegraphics{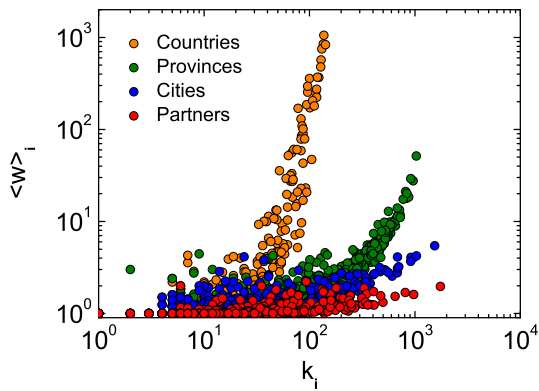}
}
\caption{ Average number of collaborations $\left < w\right >_{i}$ of each node $i$ versus the degree $k_{i}$ of the node. With red circles we represent the individual partners (19984 institutions), with blue circles the cities (6955), with green circles the provinces (1304), and with orange circles we represent the countries (154).
}
\label{fig:2}
\end{figure}

\section{Analysis of the collaboration network}

We construct the collaboration network by considering each participant in the FP as a node, and by representing the collaborations between partners as links between the nodes. Furthermore, we consider as link weight, $w_{ij}$, of the connection between nodes $i$ and $j$ the total number of collaboration projects between these two partners. In this view we can construct the network using different detail levels, depending on the information we would like to extract. This ``zooming'' in or out, is achieved by considering as nodes either the countries, provinces, cities, or the individual institutions participating in the FP. 

We first construct the network of collaboration between individual partners (institutions), and then we calculate the degree of each node. The degree $k_{i}$ of node $i$ is defined as the number of links that are connected to this specific node $i$. In fig.~\ref{fig:1}(a) is shown the probability distribution of the node degrees, $P(k)$, which is used to calculate the probability of finding a node with degree $k$, for both the FP5 and FP6 collaboration networks. We identify two different regions in the plot, as separated by a vertical dot line around the value of $k \sim 8$. In the region to the right of the dotted line, both probability distributions have a clear power law tail, $P(k)\sim k^{-\gamma}$. For the FP5 network the exponent $\gamma=1.85\pm0.03$, a value in good agreement with~\cite{bb:MendesFP5}, the maximum degree is $k_{max}=2784$, and the mean degree of the network is $<k>=26.1$. For the FP6 network the exponent $\gamma= 1.96\pm0.03$. The maximum degree of the network for FP6 data is $k_{max}=2842$, and the mean degree of the network is $<k>=43.6$. Such linear decay in the log-log plot is typical for many networks~\cite{bb:AlbertBarabasi, bb:Faloutsos, bb:ShaiCarmi, bb:AlbertJeongBarabasi, bb:PastorVespignani, bb:Garlaschelli, bb:Liljeros, bb:NewmanPnas}.
In the left region, the probability distribution for the FP5 is always larger than the corresponding one for the FP6. This means that there are more institutions in the FP5 having a smaller number of connections than in FP6. The opposite behavior is found in the right region, where the probability distribution for the FP6 is always larger than the one for the FP5, which means that the number of participants with a larger number of connections has been increased.  The large increase in the value of $< k >$ reveals that the number of collaborative partners per institution has been increased. This means that there are more links connecting the individual partners, and these links result in an enhancement of the collaboration activity between institutions over time, even though the number of signed contracts was decreased by a factor of 2, and the total number of unique partners also considerably decreased. Similar behavior, i.e. the increase of the mean degree, is found when we consider the collaboration network based on the different funding schemes, as we see in fig.~\ref{fig:1}(b). We also observe here a scale-free behavior as with the total number of projects. Therefore, we find that the different funding schemes lead to the same collaboration behavior, even though the requirements to establish collaborations are not entirely the same for each scheme.

\begin{figure}
\resizebox{1.0\columnwidth}{!}{%
\includegraphics{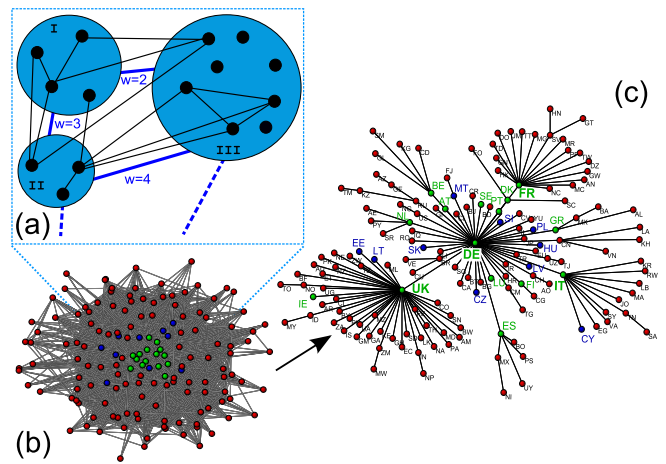}
}
\caption{(a) Pictorial representation of how the network of collaborations among countries is created. In this example with black edges we represent possible connections among institutions (small circles) and with blue edges the connections between countries (large circles). The number of connections between institutions belonging to two different countries is the weight of the connection between them. (b) The actual network of collaborations among the 154 countries participating to at least one FP6 project. (c) The Minimum Spanning Tree (MST) of the collaboration network of the 154 countries participating at least to one FP6 project. The green nodes represent the EU15 member countries, the blue nodes represent the 10 new member countries of EU, and the red nodes represent all the other countries.
}
\label{fig:3}
\end{figure}

Next, we study the dependence of the mean node weight, $\left<w\right>_{i}=\frac{1}{k}\sum_{j}w_{ij}$ , on the node degree $k$ for all detail levels that we have available (fig.~\ref{fig:2}). When we consider the network of individual participants (red dots) the mean node weight $\left<w\right>$  is almost uniform with a value in the interval between $\left<w\right>=1$ and $\left<w\right>=2$, for all nodes. Therefore, there is no traceable dependence of this value to the node degree $k_{i}$. We assume that this behavior is due to the fact that, among the large number of total projects the most frequently  appearing case is the case of only one or two collaborative projects between the individual institutions, and thus, the mean node weight is almost equal to the node degree found here. But this picture changes as we zoom out to cities, provinces, and eventually to country level. For the case of countries, as it is shown in fig.~\ref{fig:2} with orange circles, there is a tendency of nodes with high degree to have higher mean node weight $\left<w\right>$ values. For example, we find that Germany, one of the leading countries in research in Europe, has $\left<w\right> = 1053$ and $\left<k\right> = 142$. This happens because there are only a few countries which are very far from the average in terms of the projects carried out by them, as compared to the rest of the countries. Such countries (see later in fig.~\ref{fig:3}), are the hubs of the network and play a more important role in its structure. The data for cities and provinces (blue and green circles in fig.~\ref{fig:2}) behave similarly, and are located in the intermediate range of the plot.

To understand better how the aforementioned behavior occurs, we discuss the example network of fig.~\ref{fig:3}(a), where a pictorial representation of the collaboration network in different levels is shown. In this network, with black edges we represent possible connections among institutions (small black circles) and with blue edges the connections between countries (large blue circles). The number of connections between all institutions belonging to two different countries is the weight of the connection between these two countries. We see that while for the institutions the mean node weight is equal to the node degree, for countries their value is different. In this figure all three countries have $k = 2$ but for country I $\left<w\right>_{I} = 2.5$, for country II $\left<w\right>_{II} = 3.5$, and for country III $\left<w\right>_{III} = 3$. From this example we see that country II plays more important role in the network, since it has the largest average weight, even though the other two countries have larger number of institutions. This happens because one of the institutions of country II is connected to many different institutions of other countries, and thus, it increases the connection strength of the country that it belongs to.

In the following analysis we focus on the collaboration network among countries participating in the FP6, as shown in fig.~\ref{fig:3}(b). An edge connecting two nodes, $i$ and $j$, of this network represents the presence of at least one collaboration project between institutions from country $i$ with institutions in country $j$. The weight $w_{ij}$ of such an edge represents the total number of collaborations between institutions in these two countries. We transform this weight to a distance measure $d_{ij} = 1/w_{ij}$ , in such a way that the smaller the distance, the stronger the collaboration between countries. By default $d_{ij}$ is defined in the interval $\left(0,1\right]$, and it takes its maximum value when there is only one collaboration project between a pair of countries, $w_{ij} = 1$. From this network we extract the Minimum Spanning Tree (MST). In order to construct the MST we use the following algorithm. First, we start with the disconnected network, which includes only the nodes with no links between them. Next, we sort all distances from highest to lowest. Next, we start adding links to the disconnected network in increasing distance order, i.e. we connect firstly the nodes with the smallest distances. If a loop is formed, this is not allowed, and the link is discarded. This procedure stops when all nodes are connected, and no further links can be added. The MST is a special graph because it reduces the number of links of the network, since as a tree it cannot contain any loops, while it keeps all the nodes connected with a total minimum distance. This structure is a much simpler graph than the full network, but it still gives interesting information about the system. The usage of spanning trees as subnetworks that retain the most meaningful connections of the original network is an approach that enhanced our understanding in various complex systems. For example, in correlation based networks of financial markets the Minimum Spanning Tree technique~\cite{bb:Mantegna} has led to the identification of clusters of stocks that result to a meaningful taxonomy.

\begin{figure}
\resizebox{1.0\columnwidth}{!}{%
\includegraphics{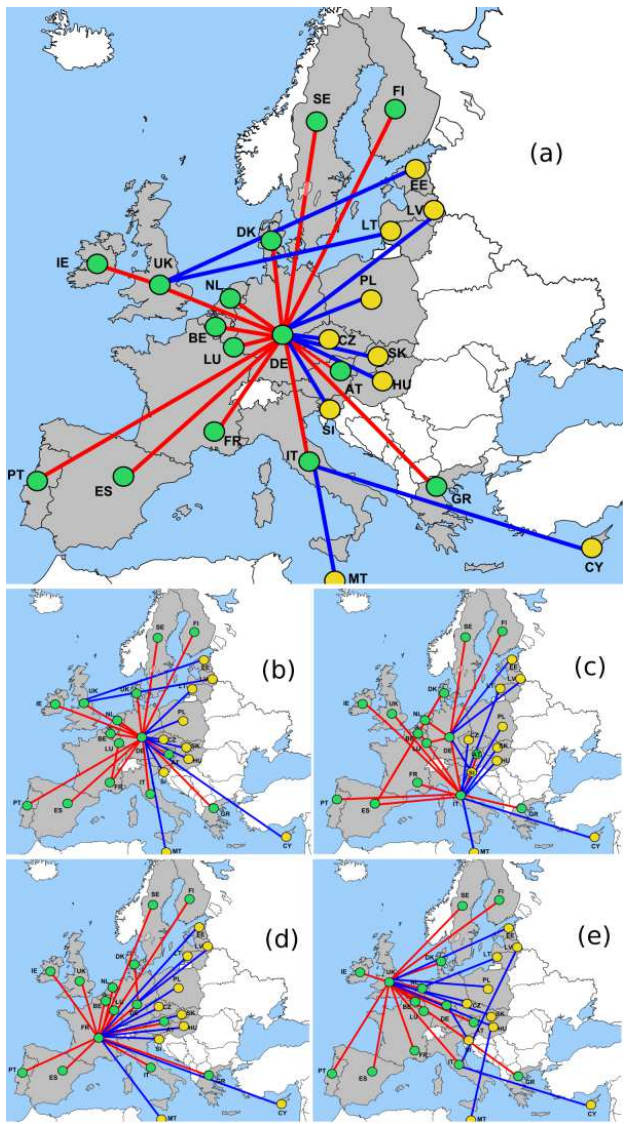}
}
\caption{(a) Minimum Spanning Tree (MST) indicative of the collaboration activity in Europe, using all thematic areas of the FP6 for all EU25 countries. MSTs indicative of the collaboration activities, for all EU25 countries, in the following thematic areas : (b) ``Nanotechnologies and Nanosciences''. (c) ``Research and Innovation''.  (d) ``Aeronautics and Space''. (e) ``Food Quality and Safety''. Colour code: with green nodes we represent the EU15 countries and with red lines we represent the strongest links between them, while with yellow nodes we represent the 10 new member countries of EU, and with blue lines we represent the strongest links of these 10 countries.
}
\label{fig:4}
\end{figure}

The MST of the FP6 collaboration network, interestingly enough, has a star-like structure around some specific countries, as it is shown in fig.~\ref{fig:3}. These countries, that act as hubs (strongly connected nodes) are Germany, United Kingdom, France, and Italy. The most connected hub, and the one with the most central role is Germany, that is located at the center of the MST in fig.~\ref{fig:3}(c). This central role becomes clearer if we examine the stronger connections only among countries belonging to the European Union. From this examination we find that almost all 15 European countries are directly linked to Germany, as it is shown in more detail in fig.~\ref{fig:4}(a), except Ireland, that is linked to the UK. When we consider the 10 new countries that joined EU in 2004 we see that 6 of them established stronger connections with Germany, but 4 with the UK and Italy.

Similar behavior is found if we calculate the MST of the collaboration network between countries for every thematic area separately, as it is shown in figs.~\ref{fig:4}(b) - \ref{fig:4}(e), where we show the MST that corresponds to collaboration activity among European countries, for four (4) different thematic areas. Namely, the thematic areas presented in figs.~\ref{fig:4}(b) - \ref{fig:4}(e) are: the ``Nanotechnologies and Nanosciences'', the ``Research and Innovation'', the ``Aeronautics and Space'', and the ``Food Quality and Safety''. In part (b) of fig.~\ref{fig:4} we see that Germany (DE) is the central hub, in (c) it is Italy (IT), in (d) it is France (FR), and in (e) it is the United Kingdom (UK). The observation that different countries act as hubs in different thematic areas is apparent and very interesting. 

The European Commission collaboration projects focus on a total of 16 different thematic areas. Due to lack of space, we show here the map with the MST only for four of them, and for clarity we focused only on the connections between the 25 EU member countries. The results for all 16 thematic areas, by taking into account links between all world countries participating in FP6 projects, are given in table~\ref{tab:1}. From this Table we can more easily locate the most central European countries (EU25) in terms of collaborations, by using the node degree of the MST. Following this methodology we find that Germany (DE) is the central hub for 62.5\% of the thematic areas, the United Kingdom (UK) 25\%, France (FR) 6.25\%, and Italy (IT) 6.25\%. 

We note here that by performing the same analysis to the FP5 collaboration network, we find the same qualitative results, i.e. Germany is again the most connected hub, but the MST of the FP5 has only 45\% common links with the MST of FP6. This is an indication that the collaboration network is not static, but it is of dynamic nature, and changing with time. 

\begin{table*}
	\caption{An indication of the ``centrality'' of each European country (EU25) for every thematic area, as it is calculated from the Minimum Spanning Tree (MST) connectivity. The thematic activity codes have the following meaning 
	A0: All Thematic Areas of FP6, A1: Citizens and governance in a knowledge-based society, A2: Support for the coordination of activities, A3: Euratom, A4: Food quality and safety, A5: Horizontal research activities involving Small-Medium Enterprises (SMEs), A6: Research infrastructures, A7: Specific measures in support of international cooperation, A8:  Information society technologies, A9: Life sciences, genomics and biotechnology for health, A10: Human resources and mobility, A11: Nanotechnologies and nanosciences, knowledge-based multifunctional materials and new production processes and devices, A12: Policy support and anticipating scientific and technological needs, A13: Research and innovation, A14: Science and society, A15: Aeronautics and space, A16: Sustainable development, global change and ecosystems.}
	\label{tab:1}
		\begin{center}
		\setlength{\tabcolsep}{2pt}
			\begin{tabular}{l ccccccccccccccccccccccccc}
			 \hline\hline
				   &AT & BE & CY & CZ & DE & DK & EE & ES & FI & FR & GR & HU & IE & IT & LT & LU & LV & MT & NL & PL & PT & SE & SI & SK & UK\\
			 \hline
				A0 & 1 & 5 & 1 & 1 & 47 & 3 & 1 & 5 & 2 & 17 & 3 & 1 & 1 & 11 & 1 & 1 & 1 & 2 & 4 & 1 & 1 & 1 & 1 & 1 & 42\\
				A1 & 3 & 9 & 1 & 1 & 5 & 1 & 1 & 2 & 1 & 5 & 1 & 1 & 1 & 6 & 1 & 1 & 1 & 1 & 1 & 1 & 1 & 1 & 1 & 1 & 41\\
				A2 & 5 & 3 & 1 & 2 & 23 & 1 & 1 & 2 & 2 & 10 & 1 & 1 & 1 & 1 & 1 & 1 & 1 & 1 & 1 & 1 & 1 & 1 & 1 & 1 & 3\\
				A3 & 1 & 1 & 1 & 3 & 19 & 1 & 1 & 1 & 1 & 10 & 3 & 1 & 1 & 3 & 1 & 1 & 1 & 0 & 1 & 1 & 1 & 1 & 1 & 1 & 4\\
				A4 & 2 & 9 & 1 & 1 & 4 & 4 & 1 & 3 & 2 & 2 & 1 & 2 & 1 & 7 & 1 & 0 & 1 & 1 & 12 & 1 & 1 & 1 & 1 & 1 & 35\\
				A5 & 1 & 1 & 1 & 1 & 15 & 1 & 1 & 8 & 1 & 2 & 2 & 1 & 1 & 9 & 1 & 0 & 1 & 1 & 1 & 2 & 1 & 1 & 1 & 1 & 9\\
				A6 & 1 & 2 & 1 & 2 & 29 & 1 & 1 & 2 & 1 & 1 & 1 & 2 & 1 & 17 & 1 & 1 & 1 & 1 & 2 & 2 & 1 & 1 & 1 & 1 & 4\\
				A7 & 2 & 3 & 1 & 1 & 6 & 1 & 1 & 8 & 1 & 10 & 2 & 2 & 1 & 14 & 1 & 1 & 1 & 1 & 3 & 1 & 1 & 3 & 1 & 1 & 41\\
				A8 & 4 & 6 & 1 & 4 & 40 & 3 & 1 & 3 & 1 & 14 & 3 & 1 & 1 & 6 & 1 & 1 & 1 & 1 & 1 & 1 & 1 & 1 & 1 & 1 & 6\\
				A9 & 1 & 3 & 1 & 1 & 41 & 2 & 1 & 3 & 1 & 3 & 1 & 1 & 1 & 5 & 1 & 1 & 1 & 0 & 2 & 1 & 1 & 2 & 1 & 1 & 19\\
				A10& 1 & 2 & 1 & 1 & 27 & 1 & 1 & 2 & 1 & 5 & 1 & 1 & 1 & 3 & 1 & 1 & 1 & 1 & 2 & 1 & 1 & 1 & 1 & 1 & 18\\
				A11& 1 & 2 & 1 & 1 & 38 & 1 & 1 & 3 & 1 & 4 & 1 & 1 & 1 & 4 & 1 & 1 & 1 & 1 & 2 & 1 & 1 & 1 & 1 & 1 & 4\\
				A12& 2 & 3 & 1 & 1 & 8 & 1 & 1 & 4 & 1 & 20 & 2 & 1 & 1 & 2 & 1 & 1 & 1 & 1 & 3 & 1 & 1 & 1 & 1 & 1 & 28\\
				A13& 4 & 2 & 1 & 1 & 9 & 1 & 1 & 5 & 1 & 2 & 4 & 1 & 1 & 20 & 1 & 1 & 1 & 1 & 1 & 1 & 2 & 2 & 1 & 1 & 1\\
				A14& 4 & 6 & 1 & 1 & 18 & 1 & 1 & 1 & 1 & 12 & 2 & 1 & 1 & 5 & 2 & 0 & 1 & 1 & 1 & 1 & 1 & 1 & 1 & 1 & 7\\
				A15& 1 & 1 & 1 & 1 & 4 & 1 & 1 & 2 & 1 & 35 & 1 & 1 & 1 & 2 & 1 & 1 & 1 & 1 & 1 & 1 & 1 & 1 & 1 & 1 & 1\\
				A16& 4 & 3 & 1 & 1 & 57 & 1 & 1 & 4 & 2 & 6 & 1 & 1 & 1 & 10 & 1 & 1 & 1 & 2 & 4 & 1 & 1 & 1 & 1 & 1 & 13\\
			 \hline
			\end{tabular}
	\end{center}
\end{table*}

\section{Discussion}

In this work we studied the evolution of scientific collaborations arising from the support of the European Commission, by analyzing the networks formed by universities, research centers, and private companies, which have joined their forces to address a particular scientific problem, under a funding scheme provided by the European Framework Programs. We did a comparative analysis between the FP5 and FP6, and we found that they are similar in their overall structure. We found that, even though the number of signed contracts, and the total number of unique partners from FP5 to FP6 has been decreased, the collaboration among institutions has increased, as it is revealed by a large increase in the mean degree of the network. 

We identified different dependence patterns of the average node weight $\left<w\right>_{i}=\frac{1}{k}\sum_{j}w_{ij}$ on the node degree $k$, as we ''zoom out'' from the level of participants, to the level of cities, provinces, and eventually to countries. This behavior showed that in terms of international collaborations, the more active the individual institutions of a country are, the more central role this country plays in the network structure. This centrality is demonstrated by using the MST of the collaboration networks. The MST analysis was able to distinguish the different roles of the hubs of the network for the different thematic areas of the FP6. The four larger countries, i.e. Germany, France, Italy and the UK, dominate by their participance in all FPs. The MST analysis showed that Germany is the principal overall hub when all fields of research are combined together. These findings can be taken into consideration by policy makers and government agencies when deciding future funding schemes.

\acknowledgments
We thank L.K. Gallos, who has contributed considerably to this work. This work was partially supported by the FP6 NEST/PATHFINDER project, DYSONET 012911, and by the Greek General Secretariat for Research and Technology of the Ministry of Development, PENED project 03ED840.

\end{document}